\documentstyle[12pt,epsf,epsfig,fleqn]{article}
\def\ga{\mathrel{\raise.3ex\hbox{$>$\kern-.75em\lower1ex\hbox{$\sim$}}}}
\def\la{\mathrel{\raise.3ex\hbox{$<$\kern-.75em\lower1ex\hbox{$\sim$}}}}
\newcommand{\beq}{\begin{equation}} 
\newcommand{\eeq}{\end{equation}} 
\newcommand{\bea}{\begin{eqnarray}} 
\newcommand{\eea}{\end{eqnarray}}
\begin{document}
\begin{titlepage}
\pagestyle{empty}
\baselineskip=21pt
\rightline{astro-ph/0409161}
\rightline{FTPI--MINN--04/32}
\vskip .5in
\begin{center}

{\large {\bf 
Parametrization of Dark Energy Equation of State }}

\end{center}   
\begin{center}   
\vskip 0.2in   
{{\bf Vinod B. Johri}$^{1,2}$}\\
\vskip 0.1in
{\it
$^1${Department of Mathematics and Astronomy, 
Lucknow University,\\ Lucknow 226007, India. E-mail:vinodjohri@hotmail.com }\\
$^2${William I. Fine Theoretical Physics Institute,
University of Minnesota, Minneapolis, MN 55455, USA}}\\

\vskip 0.2in
{\bf Abstract}
\end{center}
\baselineskip=18pt \noindent




A comparative study of various parametrizations of the dark energy equation of state
$w(z)$ and its variation with the redshift is made. Astrophysical constraints from 
'integrated tracking' are laid
down  to test the physical viability and cosmological compatibility of these parametrizations.
A critical evaluation of the 4-index Hannestad parametrization is done. It is found to be
simple and transparent in probing the evolution of the dark energy during the expansion
 history of the universe but it does not satisfy the BBN constraint on the dark energy
 density parameter for the best fit values; however it looks more compatible with the 
observations in the extended parameter space with $w<< -1$.

\baselineskip=18pt
\end{titlepage}
\section{ INTRODUCTION }

The recently discovered 16 Type Ia supernovae with the Hubble Telescope (Riess et al,2004)
provide a distinct scenario and conclusive evidence of the decelerating universe in
the past $(z>1)$ evolving into the present day accelerating universe. Thus the 
existence of dark energy,which accelerates the cosmic expansion , has been firmly
 established and the magnitude of its energy density has been precisely determined
(Wang and Tegmark,2004). The focus is now on the evolution of dark energy and its equation
 of state with the cosmic expansion. The simplest and the most natural candidate for
dark energy is the cosmological constant $\Lambda$ with a constant energy density
 $\rho_{\Lambda}$ and a fixed equation
of state  parameter $w = -1$. But the recent analysis of the SNe data 
(Alam et al 2004;Huterer et al 2004) indicates that the time varying dark energy
gives a better fit to observational data than a cosmological constant.

There are two contrasting prevalent views about the evolution of dark energy
at present. Riess(2004) and Jassal et al(2004) have argued that the current
SNe Ia observational data is inconsistent with the rapid evolution of $w(z)$.
On the other hand, Bassett et al(2004) contend that the inconsistency arises 
on account of the inadequacy of 2-index parametrizations. They claim that the
rapid variation in $w(z)$ in fact provides a better fit to the 'gold set' SNe
observations, even after including CMB and large scale structure data 
(Corasaniti et al,2004). According to our analysis, the rapidity of variation
$|\frac{dw}{dz}|$, apart from other factors, depends on the absolute value of $|w|$.
For quintessence models $-1<w< -\frac{1}{3}$, whereas $w$ has no lower bound 
for phantom models. Hence the equation of state of dark energy varies more rapidly
for phantom models with large $|w|$ than for quintessence models, other factors remaining unchanged.
The fact that the observational data (Caldwell et al,2002) including SNe and galaxy
clustering, shows a bias towards phantom models might explain why rapid variation
in $w(z)$ provides a better fit to observational data.

It is difficult to know the exact functional form of the equaion of state $w(z)$
observationally as such different parametric forms (Huterer and Turner ,2000;
Weller and Albrecht,2001;Polarski et al,2001;Linder,2003;Padmanabhan et al,2004; 
Corasaniti et al,2004; Alam et al,2003 ) and non-parametric forms (Bassett et al,2002; 
Corasaniti and Copeland,2002; Corasaniti et al,2004; Bassett et al,2004) of $w(z)$have
been used to simulate the behavior of evolving equation of state of dark energy
and comparison made with the SNe observations. The SNe observations essentially 
measure the luminosity distance $d_L(z)$ which when compared with the theoretical 
parametric values, yields the best fit values of the parameters (For details see
Padmanabhan,2003). Odintsov and Nojiri(2004) have discussed the thermodynamics of
the evolution of dark energy and have stressed that the phantom stage might be a
transient period in cosmic evolution.
 
For making a comparative study of various parametrizations and their viability, we need 
some criterion.
In our previous papers (Johri,2001,2002),we introduced the concept of 'integrated tracking'
and outlined certain astrophysical constraints to be satisfied by the dark energy fields
for realistic tracking. For example, the dark energy density parameter at nucleosynthesis 
epoch $(z = 10^{10})$ must satisfy the constraint

$(A)\quad\quad (\Omega_X)_{BBN}\;\; < \;\;0.14$

The latest analysis (Olive et al,2004) constrains

$(A)\quad\quad (\Omega_X)_{BBN}\;\;  <  \;\; 0.21$

Again at the point of transition from deceleration to acceleration phase of expansion,
a viable parametrization must satisfy the constraint (corresponding to $q=0$).

$ (B)\quad\quad (\Omega_X)_T\; = \;  -\frac{1}{3w(z_T)}$

In this paper, we have discussed the merits and demerits of four dark energy 
parametrizations, first three involving 2-parameters and the fourth (Hannestad,
2004) involving 4-parameters.  More parameters 
mean more degrees of freedom for adaptability to observations, at the same time 
more degeneracies in the determination of parameters. Hannestad parametrization
is critically examined for its effectiveness and limitations to represent the
variation of $w(z)$ in compatibility with the observations.

The outline of the paper is as follows
In section 2, we have discussed the expansion history of the universe.
Assuming a spatially flat universe,the field equations give the Hubble 
expansion and deceleration parameter in terms of function $f(z)$ which 
involves integral of the varying equation of state of dark energy.
In section 3, we have discussed 2-index parametrization models by Huterer 
and Turner(2002),Weller and Albrecht(2002), Linder(2003) and Padmanabhan 
et al (2004).
In section 4, we have critically examined Hannestad parametrization model
which involves 4 parametrs imparting four degrees of freedom to choose them.
In section 5, we use interpolation technique to study the behavior of the 
dark energy  under the assumption of slowly varying equation of state.
 In section 6, we conclude with some remarks on parametrization methods.

\section{EXPANSION HISTORY OF THE UNIVERSE }

Assuming a spatially flat  (k=0) friedmann universe, the field equations are
\begin{equation}
H^2\;\;=\;\;H_o^2[\Omega_M^o(1+z)^3 + \Omega_X^o f(z)]
\end{equation}
and
\begin{equation}
\frac{2q-1}{3}\;\;=\;\; w(z)\Omega_X
\end{equation}
where $H=\frac{\dot{a}}{a}$ is the Hubble constant, $q= - \frac{\ddot{a}}{aH^2}$
is the deceleration parameter, $w(z)=\frac{p_X}{\rho_X}$ is the equation of state of 
dark energy and $\Omega_X= 1-\Omega_M$ is the cosmic dark energy density parameter
with $\Omega_X=\Omega_X^o f(z)$,
\begin{equation}
f(z)\;\; = \;\;\exp\big(3\int_0^z{\frac{1+w(z')}{1+z'}dz'}\big)
\end{equation}
The luminosity distance is given by
\begin{equation}
d_L(z)\;\; = \;\;(1+z)\int_0^z{\frac{dz'}{H(z')}}
\end{equation}

We can test any  parametrization  by taking the best fit values and applying
constraints (A) and (B) to check its compatibilty with the 
cosmological observations. The transition redshift may also be derived from the best 
fit values.
\section{Two Index Parametrizations}

I.  Linear Redshift Parametrization of $w(z)$.
 
   (Huterer and Turner,2001; Weller and Albrecht,2002)
\begin{equation}
w(z)\;\; = \;\; w_o + w'z,\quad\quad w' =  \Big( \frac{dw}{dz}\Big)_{z=0}
\end{equation}
Inserting for $w(z)$ in Eq.(3), we get
\begin{equation}
\Omega_X\;\; = \; \Omega_X^o(1+z)^{3(1+w_o-w')}\times\exp[3w'z]
\end{equation}
It has been used (Riess et al, 2004) for probing SNe observations at $z<1$.
It diverges for large $z$. The best fit values for SNe 'gold set'(Dicus
and Repko,2004) , $w_o= -1.4, w'= 1.67, \Omega_M^o=0.3, z_T=0.39$ give
$\Omega_X(z_T)=0.63$ whereas constraint (B) demands $\Omega_X=0.445$.
This parametrization gives $+40\%$ deviation from the acceptable value 
of the dark energy density.

II.Polarski and Linder Parametrization (Polarski,2001;Linder,2003)

\begin{equation}
w(z)\;\; = \;\; w_o + w_1\frac{z}{1+z}
\end{equation}
On differentiation, $\frac{dw}{dz}\;=\; \frac{w_1}{(1+z)^2}$
It indicates rapid variation of $w(z)$ at z=0 which goes on decreasing with increasing $z$.
For large $z$, $w(z)= w_o + w_1$. 

Eq.(3) gives
\begin{equation}
\Omega_X\;\; = \;\Omega_X^o(1+z)^{3(1+w_o+w_1)}\times\exp[-\frac{3w_1z}{1+z}]
\end{equation}
Best fit values for SNe 'gold set' (Dicus and Repko,2004;Gong,2004)
$w_o=-1.6,w_1=3.3,\; \Omega_M^o = 0.3,  z_T=0.35$ give $\Omega_X^T= 0.626$
whereas contraint (B) demands $\Omega_X^T= 0.46$; as such this parametrization
gives $+36\%$ deviation from the acceptable value of dark energy density. 
Since $w_o+w_1>0$, 
$\Omega_X$ diverges for large $z$. Also the BBN constraint (A) is not satisfied.

III. Padmanabhan Parametrization(Jassal, Bagla and Padmanabhan,2004),

\begin{equation}
w(z)\;\; = \;\; w_o + \frac{w_1 z}{(1+z)^2}
 \end{equation}
For $z>>1, w(z) = w_o$. Also $\frac{dw}{dz} =\frac{w_1(1-z)}{(1+z)^3}$.
Therefore $w(z)$ increases from $z=0$ to $ z=1$ thereafter starts 
decreasing.
Eq.(3) gives
\begin{equation}
\Omega_X\;\;= \;\; \Omega_X^o(1+z)^{3(1+w_o)}\times\exp[\frac{3w_1z^2}{2(1+z)^2}]
\end{equation}
Best fit values for SNe 'gold set' $w_o=-1.9,w_1=6.6,\Omega_M^o=0.3, z_T=0.3$
give $\Omega_X(z_T)=0.583$ whereas constraint (B) demands $\Omega_X(z_T)= 0.457$
i.e. the parametrization gives a $+27\% $deviation from the acceptable value of dark energy 
density. The BBN constraint (B) is satisfied since for $z>>1$, 
$\Omega_X(z)\;\;\sim\;\;(1+z)^{3(1+w_o)}$ and $1+w_o\;<\;0$.

Therefore, Padmanabhan parametrization is compatible with astrophysical constraints 
and integrated tracking.

\section{Hannestad Parametrization}

Let us now consider Hannestad parametrization (Hannestad and Mortsell,2004) 
which involves 4-parameters.
\begin{equation}
w(z)\;\; =\;\;w_o w_1\frac{a^n+a_T^n}{w_1a^n+w_oa_T^n}\;\;         
  =\;\;\frac{1 + \big(\frac{1+z}{1+z_T}\big)^n}{w_o^{-1}+w_1^{-1}\big(\frac{1+z}{1+z_T}\big)^n}
\end{equation}

where $a=(1+z)^{-1}$, $a_T$ is the scale factor at the transition redshift $z_T$, 
$w_0$ and $w_1$ are the asymptotic values of $w(z)$ as $1+z\longrightarrow0$
 and $z\longrightarrow\infty$ respectively; n determines the rapidity of transition.

It follows that the equation of state at transition epoch is given by
\begin{equation}
w(a_T)\;\; = \;\;\frac{2w_ow_1}{w_o+w_1}
\end{equation}

and the equation of state at the present epoch $(z=0)$ is given by
\begin{equation}
w^*\;\;=\;\;\frac{w_ow_1(1+a_T^n)}{w_1 + w_oa_T^n}
\end{equation}
Out of the 4 parameters, $w_o$ and $w_1$ are taken to represent the asymptotic values
and $w^*$ and $n$ are left free; alternatively $w_1$ and $w^*$ may be given pre-assigned 
values and $w_o$ and $n$ taken as free parameters to match with the observations.

Differentiating Eq.(11) with respect to z, we get
\begin{equation}
w^{-1}\frac{dw}{dz}\;=\;\frac{(w_1-w_o)n(1+z)^{n-1}a_T^n}{[1+(a_T/a)^n][w_1+w_o(a_T/a)^n]}
\end{equation}

Hence the equation of state of dark energy varies with the redshift unless $w_1=w_o$ or
$w(z)$ takes asymptotic values $1+z\rightarrow0$ or $1+z\rightarrow\infty$.
The variation gradient $|\frac{dw}{dz}|$ varies directly as $|w|, |w_1-w_o|, n$ and inversely
as $a_T^n(1+z)^n$.

At the present epoch$(z=0)$
\begin{equation}
\frac{dw}{dz}\; =\;\frac{w_ow_1(w_1-w_o)na_T^n}{(w_1+ w_oa_T^n)^2}
\end{equation}
At transition epoch$(z=z_T)$
\begin{equation}
\frac{dw}{dz}\;\;=\;\;\frac{w_ow_1(w_1-w_o)na_T^n}{2(w_1+w_o)}
\end{equation}
Note that $\frac{dw}{dz}$ changes sign in going from $z=0$ to $z=z_T$ if 
$w_o<0,w_1<0$ and $ w_1-w_o >0$.

Inserting for $w(z)$ from Eq.(11) into Eq.(3), we get
\begin{equation}
\Omega_X(z)\;=\;\Omega_X^o(1+z)^{3(1+w_1)}\times\Big[\frac{(w_1+w_o a_T^n)(1+z)^n}
{w_1+w_oa_T^n(1+z)^n}\Big]^{3(w_1-w_o)/n}
\end{equation}

For $z>>1$

\begin{equation}
\Omega_X(z)\;\;=\;\Omega_X^o(1+z)^{3(1+w_1)}\times\big(1 + \frac{w_1}{w_o a_T^n}\big)^{3(w_o-w_1)/n}
\end{equation}

Taking the best fit values for the 'gold set' SNe (Hannestad,2004),

$w_1 =\;-0.4; w_o =\;-1.8 ; n\;=\;3.41$ with a prior $\Omega_M^o=0.38$,
we have $w(z_T)\;=\;-0.654$, by Eq.(12). 

The fourth parameter $w^*$ is still free; we can choose $w^*\;=\;-0.95$, 
a value favored by observations, then Eq.(13) yields $z_T\;=\;0.367$.
Substituting for $z_T$ in Eq.(17), we find $\Omega_X(z_T)\;=\;0.68$
whereas constraint (B) demands $\Omega_X(z_T)=\;-\frac{1}{3w(z_T)}=0.509$.
Thus Hannestad parametrization gives $+33.3\% $ deviation from the acceptable
 value of dark energy density. Again, it is obvious from Eq.(18) that the 
dark energy density parameter diverges for large values of redshift since
$1+w_1>0$ for the best fit values given above; as such constraint (B) is not
satisfied.

\section{Interpolation of w(z) for Slowly Varying Equation of State}

If we go by the analysis of Riess et al(2004) and Jassal et al(2004), 
the current SNe data is inconsistent with rapid evolution of dark energy.
Therefore, we can use 'integrating tracking' and interpolation techniques
applicable to slow time varying equation of state (Johri,Pramana,2002),
It was shown (Johri, Class. \& Quantum Gravity,2002) that the scalar fields 
with slowly varying equation of state which satisfy tracking criteria, are
 compatible with astrophysical constraints (A) and (B) outlined under 'integrated
tracking'. According to our analysis (Johri, Pramana,2002), based on integrated
tracking, we have
\begin{displaymath}
\begin{array}{lll}
 \Omega_X\;\leq\;\; 0.14, & w\simeq-0.035\;\;\;& {\rm at}\;\; z\;=\;10^{10} \\
 \Omega_X\;\;=\;\;0.66, & w\;=\;-0.77\;\;& {\rm at}\;\; z\;\;=\;0 \\
 \Omega_X\;\;=\;\;0.5, & w\;\;=\;-0.66\;\;& {\rm at}  \;\; z_T\;=\;0.414
\end{array}
\end{displaymath}
The above data is  consistent with the transition redshift 
given by Riess et al(2004) and $(\Omega_X)_{BBN}$  given by Cyburt et al(2004)
To apply Hannestad parametrization,if we choose asymptotic values $w_1= -0.035,w_o= -1$ (assuming quintessence 
dark energy),we get $w(z_T)= -0.676$ which
yields  $z_T = 0.405$ and $\Omega_X(z_T) = 0.493$. These values are remarkably close to 
the interpolated data given above.

\section{Conclusion}

We have investigated the behavior of evolving dark energy by taking parametric 
representation of the equation of state. Out of the 2-parameter models of the 
equation of state, Padmanabhan
parametrization is found to be valid and compatible with the 
astrophysical constraints over a wide range of redshift.

Bassett et al (2004) have discussed the limitations of 2-index
parametrizations. Most of them track well at low redshifts
$z\leq0.2$ but if we explore dark energy for regions $z>1$ and $ w<< -1$
(beyond quintessence models),  very rapid variation in $w(z)$ can 
be found. Higher order parametrizations are more suitable for probing
the nature of dark energy and its evolution since more parameters give
 more freedom to fit in observational data but at the same time it 
gives rise to more degeneracies in the determination of the parameters,

Hannestad 4-parameter model of the equation of state provides a well-
behaved representation of dark energy evolution over a long range of
redshift but it fails to satisfy the BBN constraint for dark energy 
density for quintessence models $w+1>0$ at nucleosynthesis epoch
around $z=10^{10}$. However, if the weak energy condition (WEC) is 
relaxed to admit phantom models $(w<-1)$, which are more favored by 
SNe and galaxy clustering observations(Caldwell et al,2003), not only
BBN constraint is satisfied but as an additional bonus, Hannestad 
parametrization admits rapid variation of the equation of state as well.
Since there is no lower bound on $w(z)$ for the phantom models, 
$\Omega_X(z)$ also decreases steeply in the early universe if $1+w_1$
is  negative in Eq.(18).

\section*{Acknowledgments}
This work was done under the DST(SERC) Project, Govt. of India.
The author gratefully acknowledges the hospitality extended by 
W. I. Fine Theoretical Physics Insitute, University of Minnesota, 
useful discussions with
Keith Olive and valuable help of Roman Zwicky 
in the compilation of this manuscript.

\end{document}